\documentclass[aps,prb,twocolumn,superscriptaddress,groupedaddress,showpacs]{revtex4-1}

\pdfoutput=1

\usepackage{graphicx,xcolor}
\usepackage{subfigure}

\usepackage[english]{babel}

\usepackage[utf8]{inputenc}
\usepackage{dsfont}
\usepackage{mathtools}
\usepackage{amssymb}
\usepackage{amsthm}
\usepackage{mathrsfs}
\usepackage{mathdots}
\usepackage{nicefrac}

\newcommand \rfrac[2]{#1/[#2]}

\usepackage{calc}
\usepackage{pgf}
\usepackage{ifthen}
\usepackage{cancel}
\usepackage{color}

\usepackage[colorlinks, linkcolor = blue, citecolor = blue, filecolor = black, urlcolor = blue, bookmarks=false]{hyperref}
\usepackage[capitalise]{cleveref}  
\usepackage[super]{nth}
\usepackage{grffile} 

\usepackage{floatflt}

\usepackage[export]{adjustbox}

\usepackage[]{natbib}

\usepackage[]{units}

\usepackage{xspace}  

\Crefname{appendix}{App.}{Apps.}
\Crefname{equation}{Eq.}{Eqs.}
\Crefname{figure}{Fig.}{Figs.}
\Crefname{section}{Sec.}{Secs.}
\Crefname{tabular}{Tab.}{Tabs.}


\newcommand\myEqBegin{\begin{equation}\begin{aligned}[b]}
\newcommand\myEqEnd{\end{aligned}\end{equation}}

\newcounter{QMD}
\newcommand \QMDraw{\ifthenelse{\equal{\arabic{QMD}}{0}}{quasi-momentum distribution \setcounter{QMD}{1}(QMD)\xspace}{QMD\xspace}}
\newcommand \QMD{\QMDraw\xspace}
\newcommand \QMDs{\QMDraw s\xspace}

\newcounter{QSQMD}
\newcommand \QSQMD{\ifthenelse{\equal{\arabic{QSQMD}}{0}}{``quasi-stationary QMD'' \setcounter{QSQMD}{1}(QSQMD)\xspace}{QSQMD\xspace}}

\newcounter{FHM}
\newcommand \FHM{\ifthenelse{\equal{\arabic{FHM}}{0}}{Fermi-Hubbard model \setcounter{FHM}{1}(FHM)\xspace}{FHM\xspace}}

\newcounter{NNNHterm}
\newcommand \NNNHterm{\ifthenelse{\equal{\arabic{NNNHterm}}{0}}{next-to-nearest-neighbor-hopping-term (NNNH)\xspace\setcounter{NNNHterm}{1}}{NNNH\xspace}}

\newcounter{BE}
\newcommand \BE{\ifthenelse{\equal{\arabic{BE}}{0}}{quantum Boltzmann-equation \setcounter{BE}{1}(QBE)\xspace}{QBE\xspace}}

\newcounter{LBE}
\newcommand \LBE{\ifthenelse{\equal{\arabic{LBE}}{0}}{linearized Boltzmann-equation \setcounter{LBE}{1}(LBE)\xspace}{LBE\xspace}}

\newcommand \smns{\!-\!}
\newcommand \spls{\!+\!}

\newcommand \kB {k_B}

\newcommand \dk[2][3] {
	\ifthenelse{#2>3}{dk_1 \,}{}
	\ifthenelse{#2>1}{dk_2 \,}{} 
	\ifthenelse{#2>0}{dk_{#1}}{dk}
	\ifthenelse{#2>2}{\, dk_4}{}
	\ifthenelse{#2>4}{\foreach \i in {5,...,#2} { \, dk_{\i} }}{}
}
\newcommand \intdk[2][3] {
	\int
	\ifthenelse{#2>1}{}{_0^1}
	\dk[#1]{#2}
}

\newcommand \omeRaw{\omega}
\newcommand \ome {\omeRaw}

\newcommand \dK {\Delta K}
\newcommand \ddK {\delta(\dK)}
\newcommand \ddKm {\delta(\dK\!+\!m)}
\newcommand \sddKm {\sum_{\mathclap{m=-1}}^1 \ddKm}
\newcommand \dE {\Delta E}
\newcommand \ddE {\delta(\dE)}

\DeclareMathOperator \const{const}
\DeclareMathOperator \sgn{sgn}
\DeclareMathOperator \tr{tr}
\DeclareMathOperator \id{id}

\newcommand \vphdagger{{\vphantom{\dagger}}}

\definecolor{marker}{rgb}{.7,1,1}

\newcommand \kF { k_{\text{F}} }

\newcommand \fdnude{f}

\newcommand \fd {\fdnude}
\newcommand \fdm {\fd{\hspace{-1.2pt}}}

\newcommand \lam {\lambda}

\newcommand \tlam {\tilde{\lam}}
\newcommand \EF {\text{\protect\raisebox{1.8pt}{$\chi$}}}

\newcommand \emu[1] {e^{\ifthenelse{\equal{#1}{+}}{}{#1} \beta \mu}}

\newcommand \tk {\tilde{k}}
\newcommand \tkf {\tilde{k}_2(k_1,k_3)}
\newcommand \Nk {N_k}
\newcommand \Mk {M_k}
\newcommand \Ni {N_I}

\newcommand \LcRAW {\hat{\mathcal{L}}}
\renewcommand \LcRAW {{\mathcal{L}}}
\newcommand \Lc{\LcRAW}

\newcommand \Lk{\tilde{\mathcal{L}}}

\newcommand \phiv[1][n]{\tilde{\phi}_{#1}}

\newcommand \epsX {\varepsilon}

\newcommand \Ham {\hat{H}}
\newcommand \Hkin {\hat{T}}
\newcommand \nGGE {n_{\text{GGE}}}
\newcommand \ZGGE {Z_{\text{GGE}}}
\newcommand \operatorEF{\hat{Q}}
\newcommand \Iiii {\operatorEF_3}
\newcommand \n {\hat{n}}
\newcommand \ktot {\hat{K}}

\newcommand \NR {\hat{N}_{\text{R}}}
\newcommand \NL {\hat{N}_{\text{L}}}

\newcommand \order {\mathcal{O}}

\newcommand \fscalprod[2]{\langle #1, #2 \rangle_{\findex}}
\newcommand \fnorm[1]{\| #1 \|_{\findex}}
\renewcommand \fscalprod[2]{\langle #1, #2 \rangle}
\renewcommand \fnorm[1]{\| #1 \|}

\newcommand \densmat{\hat{\rho}}
\newcommand \expect[2][t]{\tr\bigl\{\densmat(#1) \, #2 \bigr\}}
\newcommand \expectShort[2][t]{\langle #2 \rangle_{\text{\raisebox{-2pt}{$#1$}}}}
\newcommand \expectShortScript[2][t]{\langle #2 \rangle_{\text{\raisebox{-1pt}{$#1$}}}}

\newcommand \JNN {J}
\newcommand \JNNN {J'\hspace{-2pt}}
\newcommand \up {\uparrow}
\newcommand \down {\downarrow}

\newcommand \om {\omega}
\newcommand \disprel [1][k]{\om(#1)}

\newcommand \groupvelRAW{\om\hspace{0.6pt}'\hspace{-.8pt}}
\newcommand \groupvel [1][k]{\groupvelRAW(#1)}

\newcommand \gamX {\gamma}  

\newcommand \cRaw {\hat{c}}
\newcommand \cd {\cRaw^{\dagger}}
\newcommand \cw {\cRaw^{\vphdagger}}

\newcommand \BEprefactorEnum{\gamX^2}
\newcommand \BEprefactorDenom{t_0}
\newcommand \BEprefactor{\frac{\BEprefactorEnum}{\BEprefactorDenom}}
\newcommand \BEprefactorLine{\BEprefactorEnum/\BEprefactorDenom}

\newcommand \SetEpsToTwoJNNNOverJNN {0}
\ifthenelse{\equal{\SetEpsToTwoJNNNOverJNN}{1}}{
	\gdef \epsDef{\epsX:=2\,\JNNN/\JNN}
	
	\gdef \NNNHprefactor{\tfrac{\epsX}{2}}
	\gdef \epsInteractionChannelNumberTransition {\tfrac12}
}{
	\gdef \epsDef{\epsX:=\JNNN/\JNN}
	
	\gdef \NNNHprefactor{\epsX}
	\gdef \epsInteractionChannelNumberTransition {\tfrac14}
}

\newcommand \Icoll { \mathcal{I}_{\text{coll}}}
\renewcommand \Icoll { \mathcal{I} }

\newcommand \EFapprox {\tilde{\EF}_3}

\newcommand \JE {\hat{J}_E}

\newcommand \JNE {\hat{J}_{N,E}}
\newcommand \jN {j_N}
\newcommand \jE {j_E}

\newcommand \jNE {j_{N,E}}

\newcommand \jNl {\hat{\jmath}_{l}^{N}}
\newcommand \jEl {\hat{\jmath}_{l}^{E}}

\newcommand \jNEl {\hat{\jmath}_{l}^{N,E}}

\newcommand \halfWidth {0.48\textwidth}

\def \FigThirdEFEpsRAW{0.01}
\ifthenelse{\equal{\SetEpsToTwoJNNNOverJNN}{1}}{
	\gdef \FigThirdEFEps{\FigThirdEFEpsRAW}
}{
	\pgfmathparse{0.5*\FigThirdEFEpsRAW}
	\pgfmathmultiply{0.5}{\FigThirdEFEpsRAW}
	\gdef \FigThirdEFEps{0.005}
	\let \FigThirdEFEpsResult\pgfmathresult
	\gdef \FigThirdEFEps{\pgfmathprintnumber[precision=1]{\FigThirdEFEpsResult}}
}

\newcommand\putFigureEigenfunctionIII{
\begin{figure}
\includegraphics[width=0.48\textwidth,page=1,trim={8 10 2 0},clip]{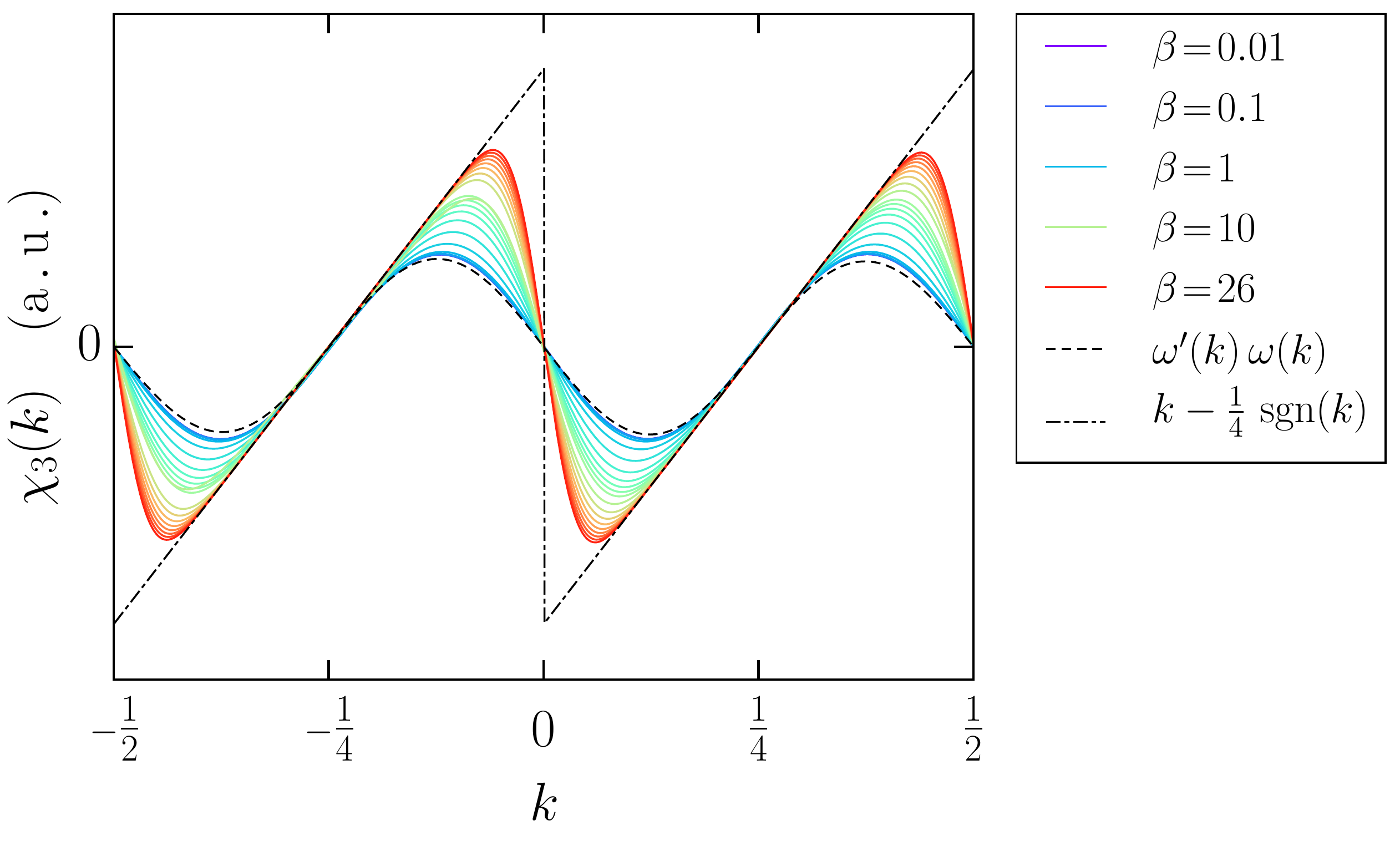}
\caption{\label{fig: 3rd EVec} 
\nth{3}  eigenfunction $\chi_3(k)$ (arbitrary vertical scale) for different final temperatures. The curves are normalized such that the slope at $k  = \frac{1}{4}$ is always the same. The relative strength of the next-to-nearest-neighbor hopping is set to $\epsX=5\cdot 10^{-3}$.}
\end{figure}
}

\newcommand\FigEVecsIVHeight{133pt}
\newcommand\putFigureEigenfunctionsIVToVI{
\begin{figure*}
\includegraphics[height=\FigEVecsIVHeight,page=4,trim={7 10 122 0},clip]{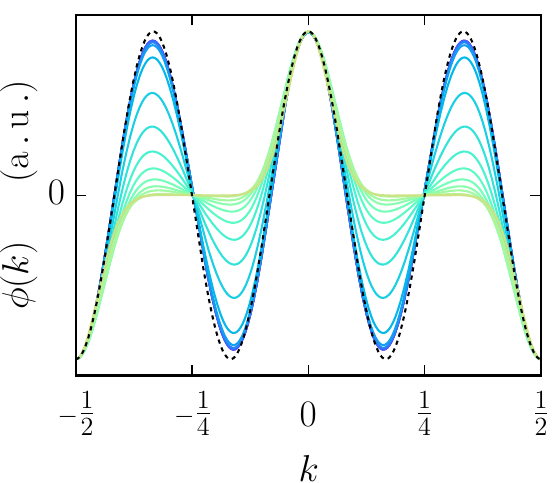}
\includegraphics[height=\FigEVecsIVHeight,page=5,trim={25 10 122 0},clip]{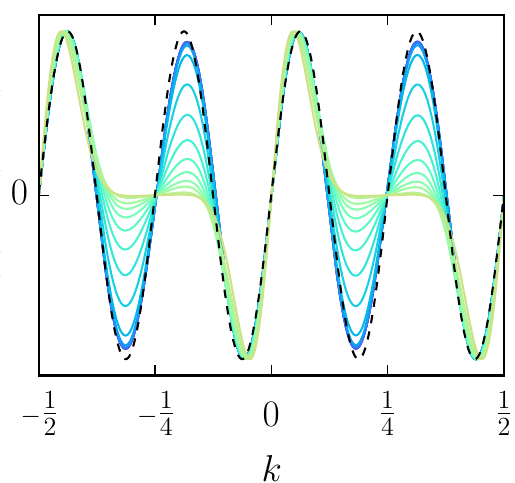}
\includegraphics[height=\FigEVecsIVHeight,page=6,trim={25 10 2 0},clip]{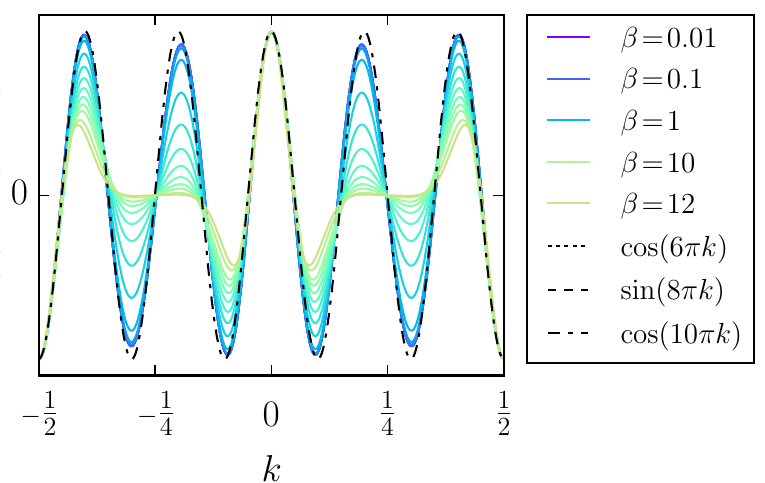}
\caption{\label{fig: Evecs 4 5 6} From left to right: \nth{4}, \nth{5} and \nth{6} eigenfunction (arbitrary vertical scale) for $\epsX=\FigThirdEFEps$ and $\beta$ ranging from $0.01$ to $12$. All eigenfunctions satisfy the approximate symmetry given by \cref{equ: Spohn-symm}. 
The curves $\sin(2\pi(n-1)k)$ describe the high
temperature limit with very good accuracy.}
\end{figure*}
}

\newcommand\FigEValsEpsHeight{170pt}
\newcommand\putFigureEigenvalueIIIEpsDependence{
\begin{figure*}
\begin{center}
\includegraphics[height=\FigEValsEpsHeight,valign=t,trim={2 7 100 0},clip]{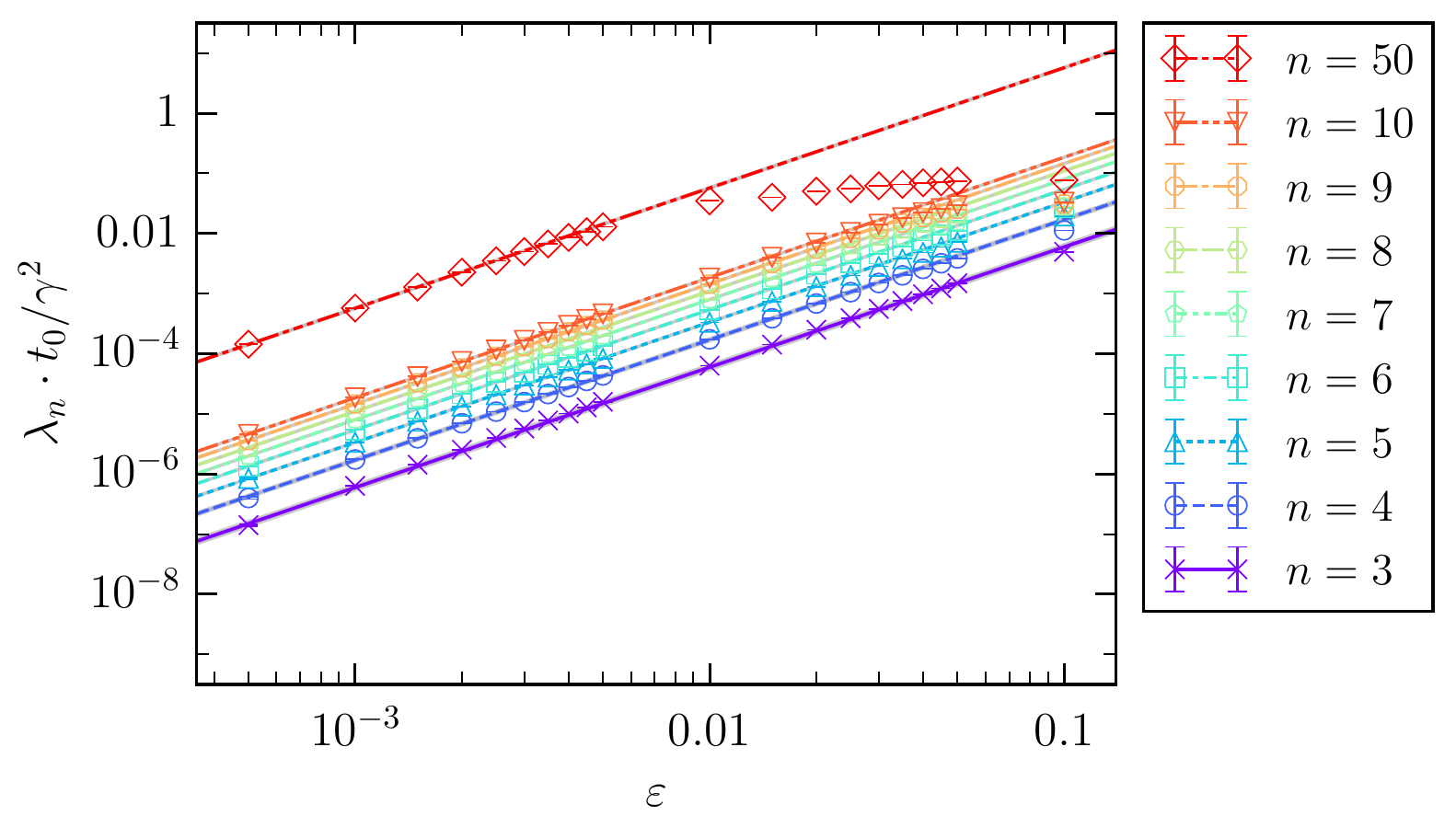}
\includegraphics[height=\FigEValsEpsHeight,valign=t,trim={60 7 2 0},clip]{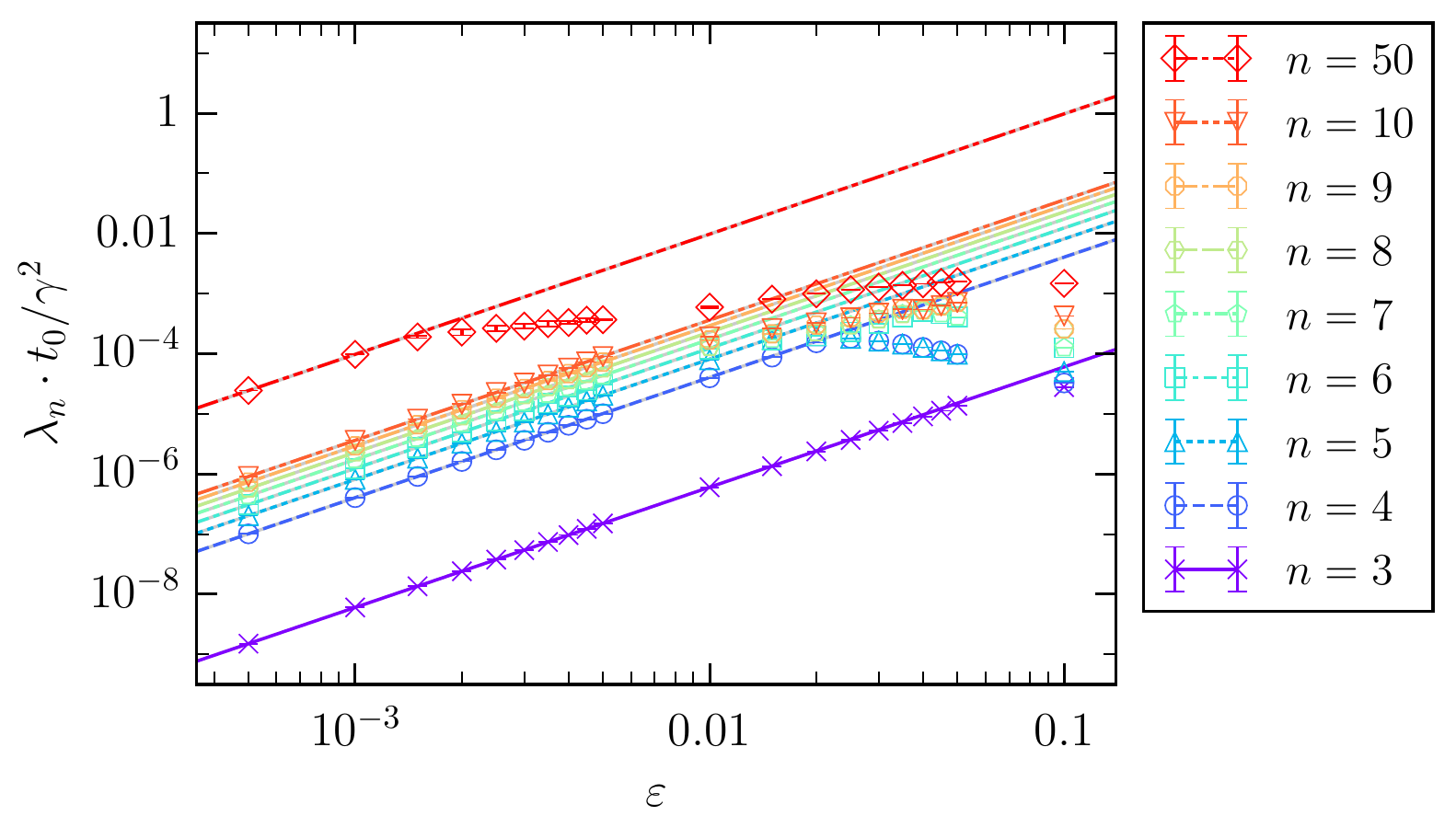}
\caption{\label{fig: JNNN dependence}
Rescaled eigenvalues $\lambda_n t_0/\gamma^2$ as a function of the relative strength of the next-to-nearest-neighbor hopping  $\epsX=J'/J$. Left plot $\beta= 0.1$, right plot $\beta = 10$. The lines are fits to quadratic behavior, $\lambda_n\propto \epsX^2$.}
\end{center}
\end{figure*}
}

\newcommand\putFigureEigenvalueIIITempertureDependence{
	\begin{figure}
	\includegraphics[width=\halfWidth,trim={0 45 0 0},clip]  {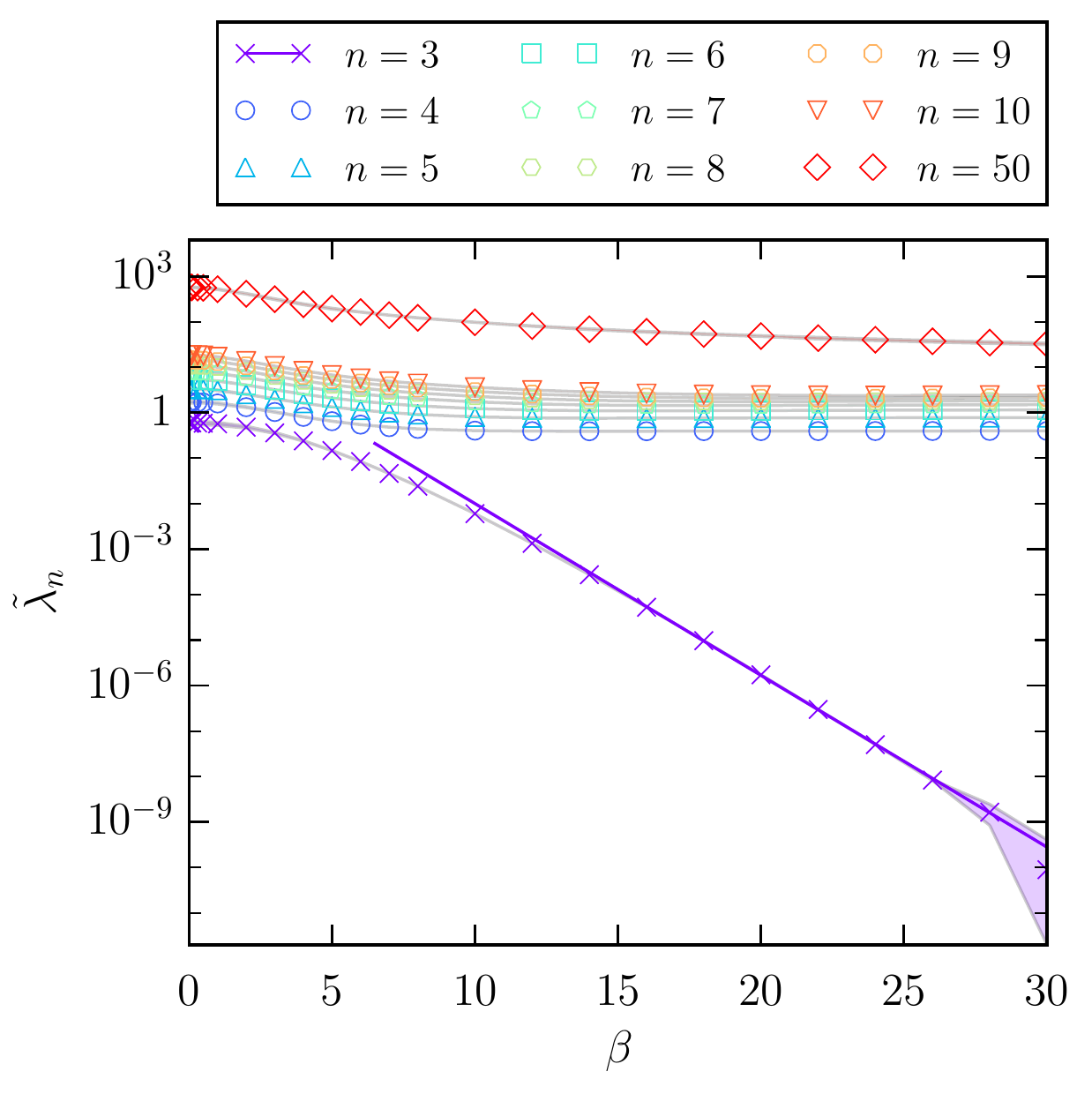}
	\includegraphics[width=\halfWidth,trim={0 45 0 76},clip] {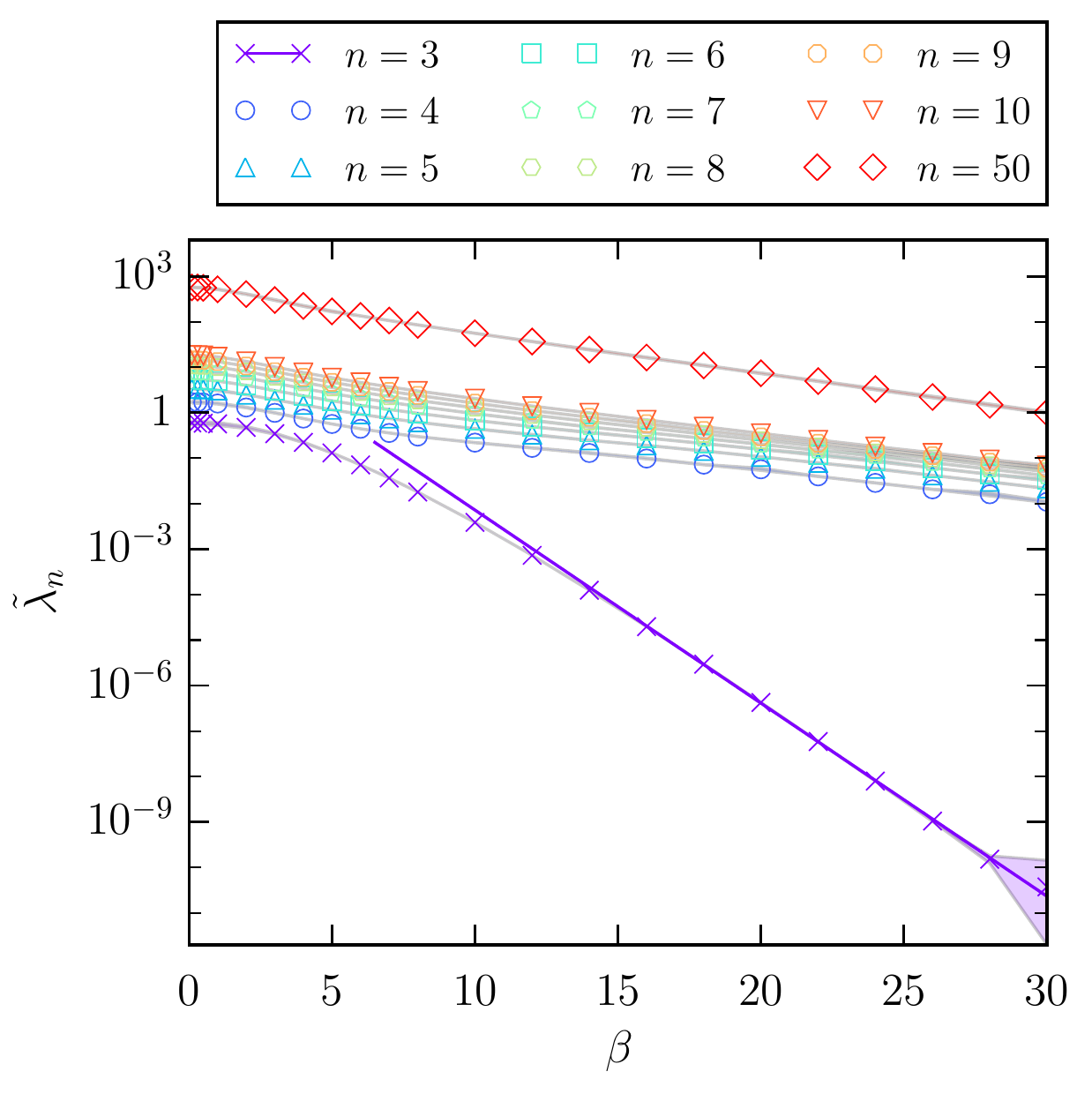}
	\includegraphics[width=\halfWidth,trim={0 0 0 76},clip]  {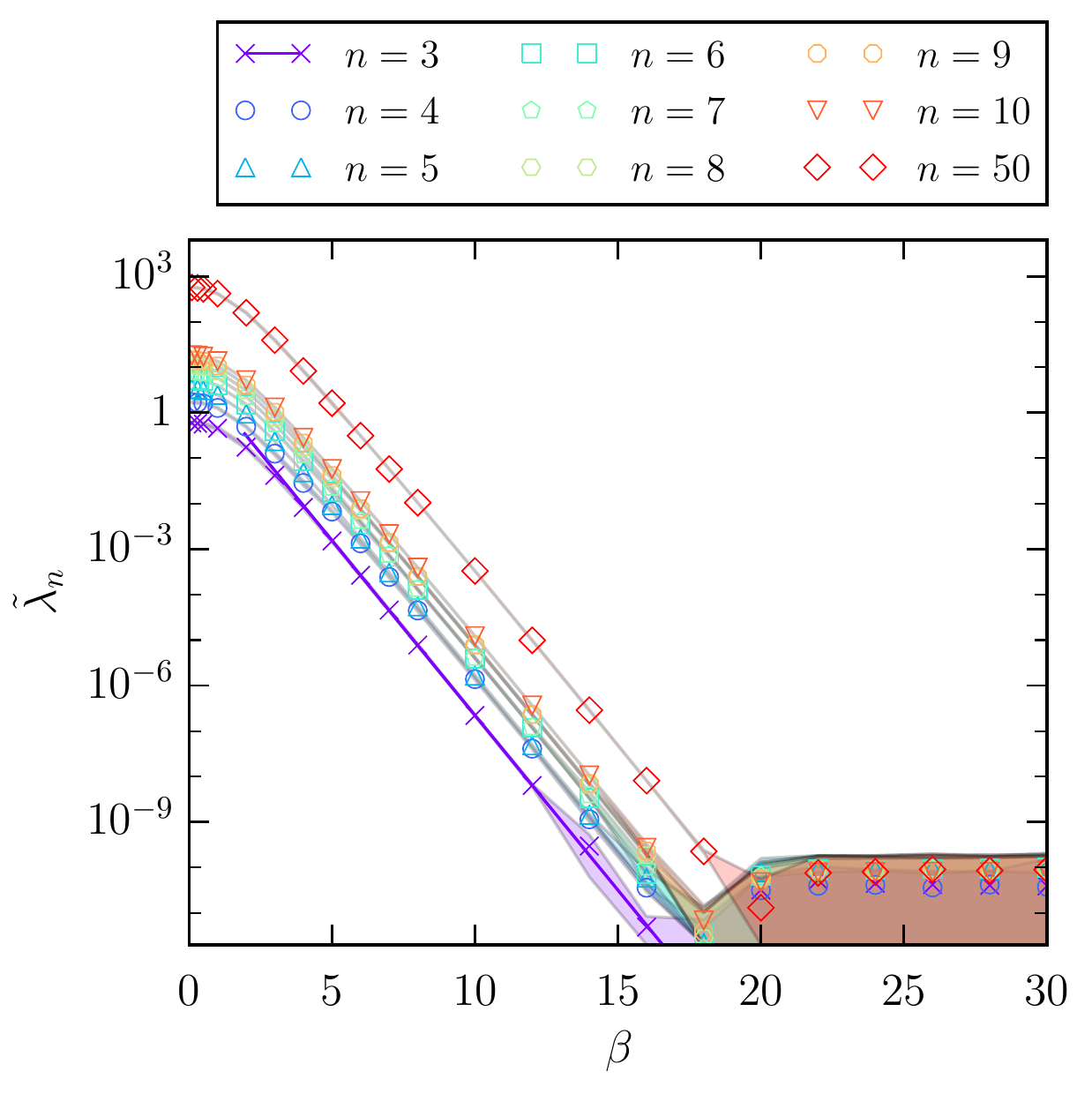}
	\\[-3pt]
	\caption{\label{fig: beta dependence} $\tlam_n(\beta)$ for $\mu=0,0.1,0.9$ from top to bottom as a function of final inverse temperature $\beta$. $\mu=0$ corresponds to half filling. The first non-zero eigenvalue $\tilde\lambda_3(\beta)$ decreases
	exponentially as a function of inverse temperature for all fillings. The higher eigenvalues ($n\geq 4$) also decay
	exponentially away from half filling, but are asymptotically constant for $T\rightarrow 0$ at half filling.}
	\end{figure}
}

\newcommand\putFigureTL{
\begin{figure*}
\includegraphics[trim={3 3 3 3},clip]{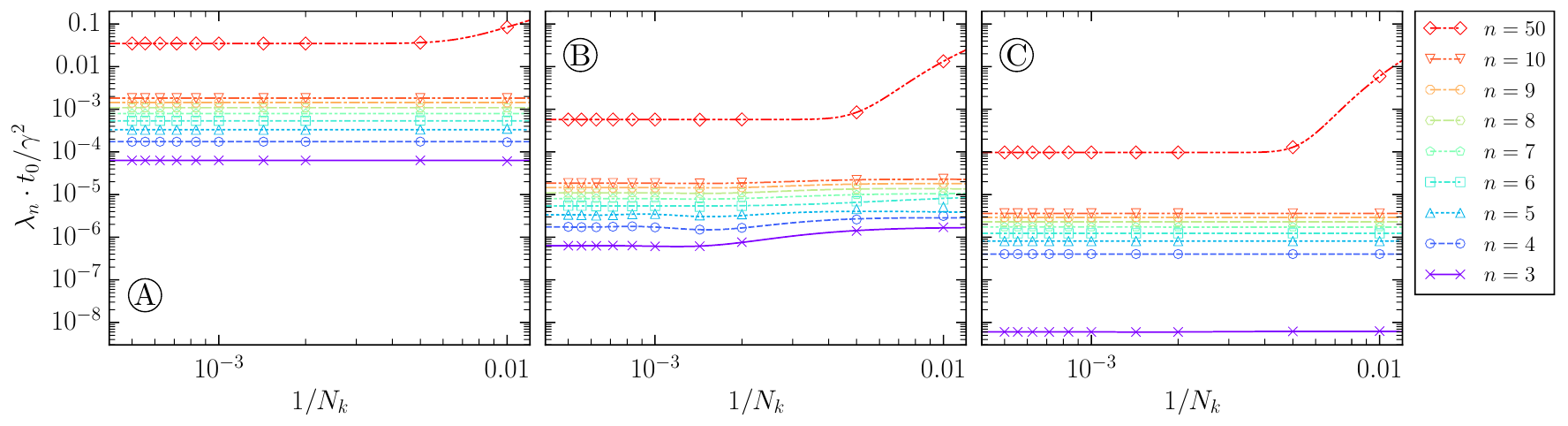}%
\caption{\label{fig: TL} We show the interpolation to $\Nk\rightarrow \infty$ for $\mu=0$. The symbols are the data for $\beta=0.1 \land \epsX=0.01$ (A). $\beta=0.1 \land \epsX=0.001$ (B) and $\beta=10 \land \epsX=0.001$ (C). The largest eigenvalue is $\lam_{50}$. The lower eigenvalues are $\lam_3$ to $\lam_{10}$ from bottom to top.  The areas around the straight lines mark the error of the fit.}
\end{figure*}
}

\newcommand\putFigureUmklapp{
\begin{figure}
\begin{center}
\includegraphics[width=\halfWidth,trim={20 0 5 0},clip]{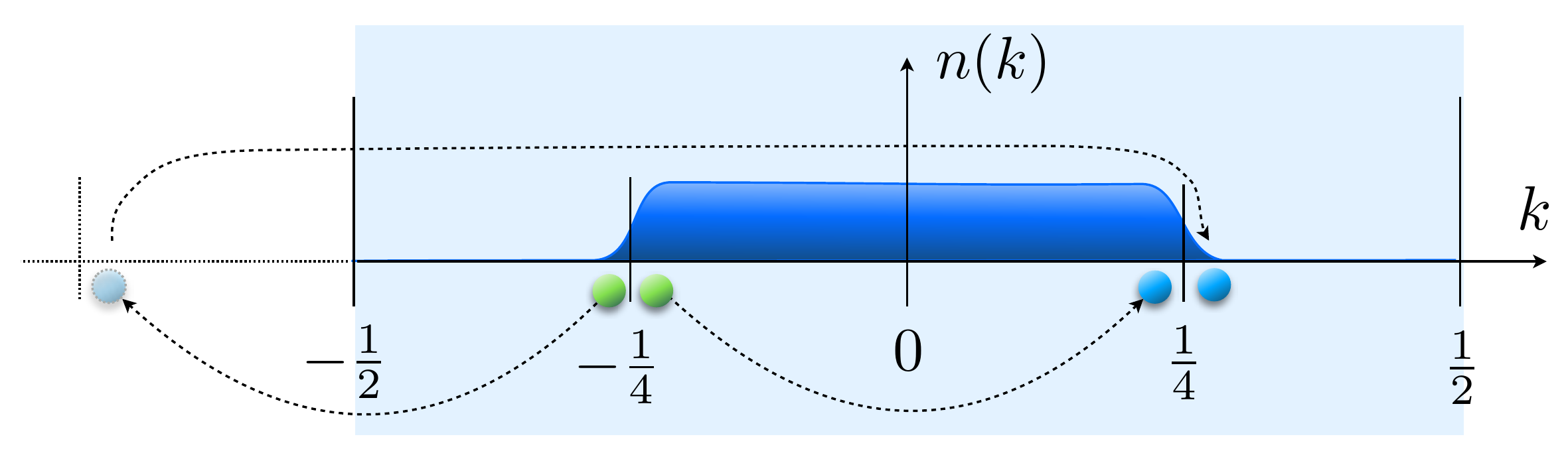}
\caption{\label{fig: Umklapp} Umklapp process at half filling near the Fermi edge $\kF\approx \frac14$.}
\end{center}
\end{figure}
}

\newcommand \epsNRaw {0.1}
\ifthenelse{\equal{\SetEpsToTwoJNNNOverJNN}{1}}{
	\gdef \epsN {\epsNRaw}
	\gdef \epsNWrite{\epsNRaw}
}{
	\pgfmathmultiply{0.5}{\epsNRaw}
	\let \epsN\pgfmathresult
	\gdef \epsNWrite{\pgfmathprintnumber[precision=2,fixed]{\epsN}}
}
\newcommand\putFigurekjs{
\begin{figure}[t]
\includegraphics[width=\halfWidth]{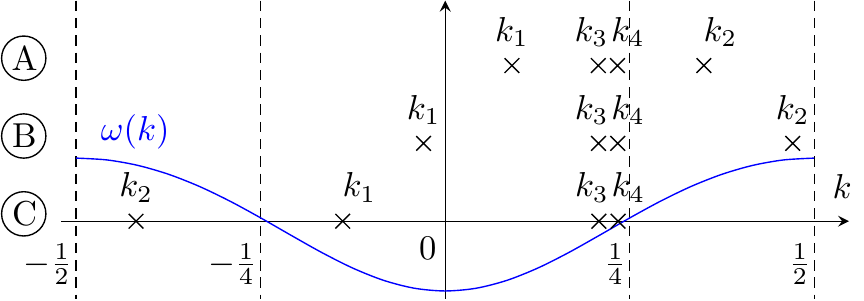}
\caption{\label{fig: LLRR} The crosses mark the momenta of forward (A), backward (B) and Umklapp (C) scattering processes for $\epsX=\epsNWrite$ conserving momentum and energy. There is only a small window allowed for backscattering processes.}%
\end{figure}
}

\newcommand\putFigureEigenvaluesTest{
	\begin{figure}
	\includegraphics[width=\halfWidth,trim={4 8 0 4},clip]{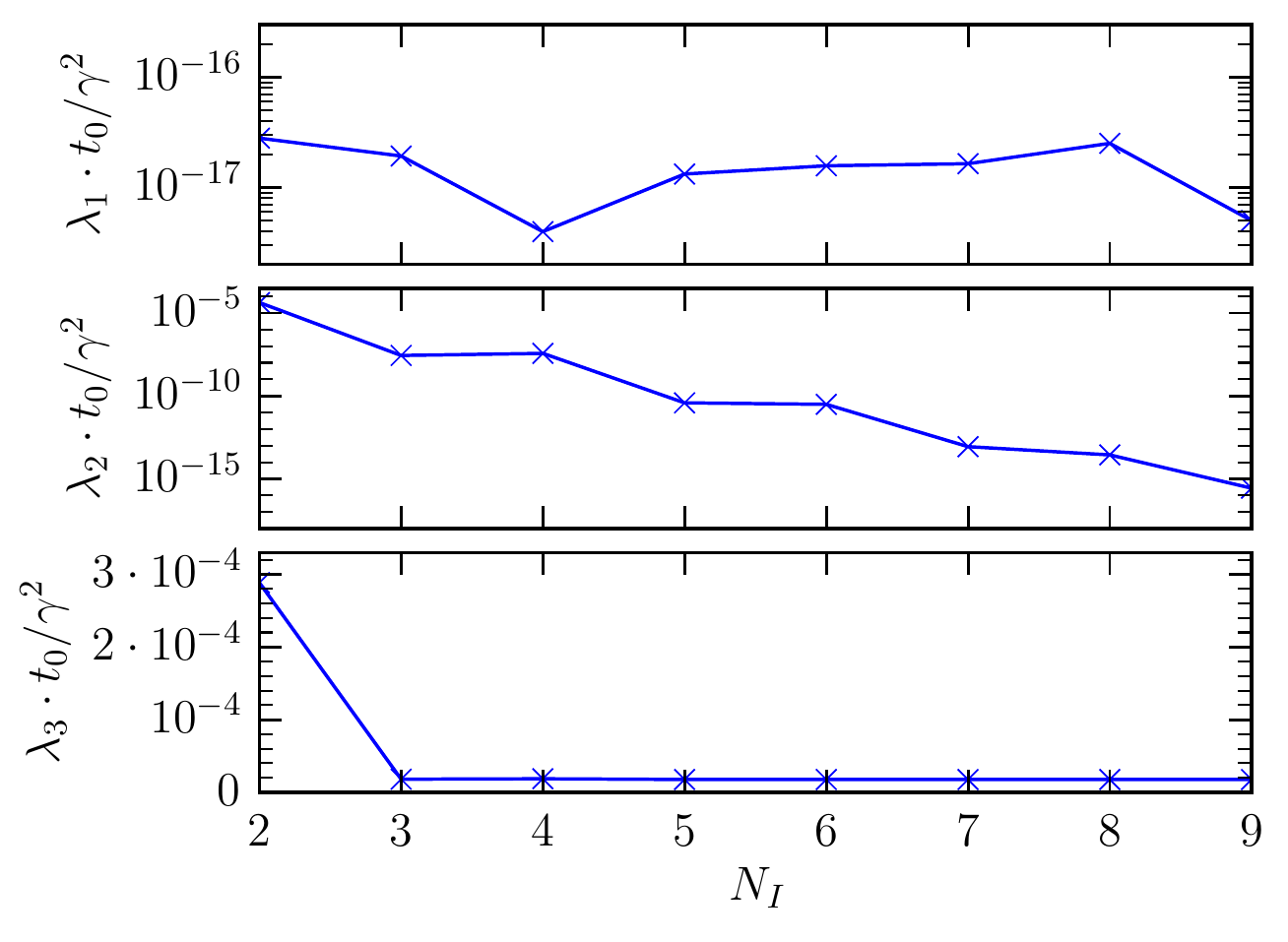}
	\caption{\label{fig: N-point-interpolation}\nth{1}, \nth{2} and \nth{3} eigenvalue for different N-point-interpolations $N_I$ for a reasonable set of parameters $\mu$, $\beta$, $\epsX$, $\Nk$.}
	\end{figure}
}

\usepackage[markup=nocolor]{changes}

\begin{document}

\hypersetup{
    colorlinks=true,
    linkcolor=blue,
    citecolor=blue,
    filecolor=black,
    urlcolor=blue,
}

\title{Thermalization rates in the one dimensional Hubbard model with next-to-nearest neighbor hopping}
\author{Fabian~R.~A.~Biebl}
\author{Stefan Kehrein}
\affiliation{Institut für theoretische Physik, Georg-August-Universität Göttingen, D-37077 Göttingen, Germany}

\date{\today}

\begin{abstract}
We consider a fermionic Hubbard chain with an additional next-to-nearest neighbor hopping term. We study the thermalization rates of the quasi-momentum distribution function within a quantum Boltzmann equation approach. We find that the thermalization rates are proportional to the square of the 
next-to-nearest neighbor hopping: Even weak next-to-nearest neighbor hopping in addition to 
nearest neighbor hopping leads to thermalization in a two-particle scattering  quantum Boltzmann equation in one dimension. 
We also investigate the temperature dependence of the thermalization rates, which away from half filling become exponentially small for small temperature of the final thermalized distribution.
\end{abstract}

\maketitle

\setcounter{QMD}{0}
\setcounter{QSQMD}{0}
\setcounter{FHM}{0}
\setcounter{NNNHterm}{0}
\setcounter{BE}{0}
\setcounter{LBE}{0}


\colorlet{darkgreen}{green!70!black}

\section{Introduction}

Understanding thermalization in quantum systems is essential to determine whether an experimental setup can be described by equilibrium concepts. Experimentally, this question becomes particularly relevant in cold atomic gases where unitary time evolution
of closed quantum many body systems can be observed because the decohering effect of the environment
is negligible (or at least under control) for the relevant time scales
\cite{Greiner2002, Kinoshita2006, Hofferberth2007,Bloch2008,Schreiber2015,Langen2015}.  
The ground breaking experiment of Kinoshita et al. \cite{Kinoshita2006} was the starting point for an ongoing
theoretical effort to understand thermalization of low dimensional quantum many body systems. Their work
considered the nonequilibrium dynamics of a 1d Bose gas with point-like interaction. Leaving aside heating and loss 
effects, they did not observe thermalization on the longest time scales accessible in their 1d experiment, while they
reported rapid thermalization for the 3d equivalent of their system. So one key theoretical question is
to understand this difference between the thermalization behavior of one dimensional and higher dimensional 
quantum systems.

A theoretical investigation of this question first requires a definition of what one means by thermalization.
Obviously, under unitary time evolution a pure state always remains a pure state and never becomes a mixed
state as employed to describe a thermal ensemble.

Therefore a more useful definition of thermalization is that the expectation values of an experimentally relevant set of observables are described by thermal values. This is the definition used in our work and we will show thermalization in this sense with respect to a certain set of observables, namely the momentum distribution function.

Theoretical research in the past decade has revealed
different thermalization behavior of integrable and non-integrable systems. While the notion of integrability in quantum systems is not uniquely  defined \citep{Caux2011}, these differences do not seem to play a role with respect to thermalization:
The long-time limit of an initial state can be described (with respect to a relevant set of observables) as a generalized Gibbs
ensemble that takes into account the expectation values of the conserved quantities of the integrable model \cite{Rigol2007}.
On the other hand, in non-integrable systems it has been
shown that even a single eigenstate can be typical for an entire thermal ensemble in the sense that expectation values
of few body observables are indistinguishable. This important observation 
regarding the foundations of quantum
statistical mechanics is called eigenstate thermalization hypothesis (ETH) \cite{Deutsch1991,Srednicki1994,Rigol2008}.
 Hence the question arises what happens at the transition from integrability to non-integrability. We will address this issue for a specific one-dimensional quantum system which is of paradigmatic importance for condensed-matter physics, namely the Hubbard model.
 
Studying the thermalization dynamics in 1d quantum systems explicitly is very challenging. For weak quantum
quenches in higher spatial dimensions one generically expects three distinct time regimes: an initial buildup
of quasiparticles, a prethermalized \citep{Eckstein2009,Gring2012,Fagotti2014,Bertini2015a} time regime (having non-thermal quasi-stationary states) and a long time thermalization described by a \BE. This
picture has been established by studying the quench dynamics of the Hubbard model for $d>2$ dimensions both analytically and 
numerically \cite{Moeckel2008,Eckstein2009}. In one dimension the general consensus is that
the Boltzmann dynamics is ineffective for two particle scattering processes due to the simultaneous conservation
of single particle energies and momenta. Therefore the thermalization 
time scale in one dimension is expected to be much longer (for example via multi particle scattering processes). Putting it otherwise, the prethermalized time regime will extend to much longer times. 

This behavior is difficult to investigate numerically. Methods like t-DMRG are limited to not too large times due to
the entanglement growth\cite{Manmana2007}, and exact diagonalization methods are intrinsically limited to small finite systems and require
an extrapolation to infinite system size. A noteable exception is a recent paper by Bertini et al. \cite{Bertini2015}, which uses a combination of numerical and analytical methods to show how
the prethermalized regime evolves towards thermal equilibrium after a quench in a dimerized 1d model of spinless fermions. 
Similarly, in our work we want to contribute to understanding the thermalization behavior of 1d systems by giving an explicit estimate for the thermalization rate based on a \BE approach in the 1d fermionic Hubbard model. Specifically, we investigate the role of an additional \NNNHterm which is tuned by the prefactor $\JNNN$
within a \BE approximation. While one sometimes finds the assertion
that there is no thermalization from a Boltzmann equation with 2-particle scattering in one dimension, we find
that this is not true in our model. This observation was already made by F{\"u}rst et al. \cite{Fuerst2012,Fuerst2013} 
and we elaborate on this initial finding
systematically in this paper by deriving all thermalization rates within a linear approximation. Only for
the case of nearest neighbor hopping only without next-to-nearest neighbor hopping does the
system not thermalize, any  nonvanishing next-to-nearest neighbor hopping $\JNNN\neq 0$ leads to thermalization in the long time limit.

The article is structured as follows. In \cref{sec:model} we introduce our model, the 1d fermionic Hubbard model (FHM) with a next-to-nearest neighbor hopping term (NNNH), and our method, the quantum Boltzmann equation approach (QBE). We will use a linearized Boltzmann equation to find the relaxation times. Furthermore we will comment on the conserved \QMDraw\cite{Fuerst2012} in the standard FHM.  \cref{sec: NumResults} is devoted to our results and the conclusions are summed up in \cref{sec: Conclusions}. \cref{sec: Stationarity of nqs} shows the stationarity of certain \QMDs. We explain our numerics in \cref{sec: NumManual}. \cref{sec: GGE} is about constructing  a generalized Gibbs ensemble (GGE) at low temperatures.
%

\newcommand \HamFHM{\Ham_{\text{FHM}}}
\newcommand \HamNNNH{\Ham_{\text{NNNH}}}

\section{Model and method}
\label{sec:model}

\subsection{Model}
\label{sec: Model}

We consider the \FHM with an additional \NNNHterm,
\myEqBegin
	&\Ham = \HamFHM + \HamNNNH
	\\ &\HamFHM = -\JNN \sum_{\mathclap{\begin{smallmatrix} l\in \mathds Z \\[1pt] \sigma\in\{\up,\down\} \end{smallmatrix}}} 
	\cd_{l\sigma} \, \cw_{l+1,\sigma} + H.c.
		+ U \sum_{l\in \mathds Z} \n_{l\up} \n_{l\down}
	\\ &\HamNNNH = - \JNNN \sum_{\mathclap{\begin{smallmatrix} l\in \mathds Z \\[1pt] \sigma\in\{\up,\down\} \end{smallmatrix}}} \cd_{l\sigma} \, \cw_{l+2,\sigma} +  H.c.
\myEqEnd
Measuring energies in units of the hopping $J$, we define dimensionless parameters $\epsDef$ and $\gamX:=U/J$. $T$ will denote the final temperature of the thermalized state, which is therefore determined by the energy of the initial state. Obviously $T=0$ implies that the initial state is the ground state. 

The dispersion relation of our model is
\myEqBegin
	\omega(k) = -\cos(2\pi k) - \NNNHprefactor \cos(4 \pi k)
	\label{equ: disprel}
	.
\myEqEnd
It is measured in units of $2J$, such that the kinetic energy is $\Hkin = 2J \int \dk0 \ \omega(k) \ \hat{n}(k)$ and the dimensionless inverse temperature is $\beta:=2J/\kB T$.

\subsection{Boltzmann-Equation}
\label{sec: Method}

A \BE describes the long time behavior of the \QMDraw\cite{Erdoes2004, Fuerst2012, Fuerst2013, Fuerst2013a}
\myEqBegin
	n_\sigma(k,t)
	= \expectShort{\n_\sigma(k)}
	= \sum_{ll'} e^{i k (l-l')} \, \expectShort{ \cd_{l\sigma} \, \cw_{l'\sigma} }
	.
\myEqEnd
Here we defined the expectation value $\expectShortScript{\hat A}:=\expect{\hat A}$. We assume the initial state to satisfy restricted quasi-freeness. This means that there is a (approximate) Wick theorem for the 4-point and 6-point functions. The \BE is valid on kinetic time scales\citep{Fuerst2013a}, i.e. times of $\order(1/U^2)$.
We operate under the normal assumption that the Boltzmann-description is still valid on longer times. 

This time evolution was previously investigated by F\"urst et al \citep{Fuerst2012,Fuerst2013} for some initial states. They found that for $\epsX=0$, the system runs into a non-thermal stationary state. However, for $\epsX\neq0$ they have seen that their initial states thermalize. The thermalization times they found for small $\epsX\neq0$ were much larger than the relaxation times for the $\epsX=0$ case.

The \BE is
\myEqBegin
	\dot{n}(k,t) = \Icoll[n](k,t)
	\label{equ: Boltzmann-equation}
	\ ,
\myEqEnd
The collision term $\Icoll$ is a non-linear operator that depends on the \QMD $n(k,t)$. Thus $\Icoll[n]$ is a function of $k$ and $t$.
We use the \BE of \citet{Fuerst2012} for the 1d FHM. 
We restrict ourselves to the spin-symmetric case in which $\up$- and $\down$-spin fermions have the same \QMD,
\myEqBegin
	n(k,t)=n_\sigma(k,t)
	.
\myEqEnd
In the spin symmetric case we can also assume
\myEqBegin
	\expectShort{ \cd_{\sigma k} \cw_{\sigma' k'} } \propto \delta_{\sigma \sigma'}
	\label{equ: delta spin}
	.
\myEqEnd
We obtain the collision term
\begin{gather}
	\begin{flalign}
	\ \Icoll[n]_1 &= \BEprefactor \, \int\limits_{\mathclap{\qquad[-\frac12,\frac12]^3}} dk_2 \, dk_3 \, dk_4 \, \sddKm \, \ddE
	&\nonumber
	\end{flalign}
	\nonumber \\
	\qquad \times \bigl[(1 \smns n_1)(1 \smns n_2)n_3n_4 - n_1n_2(1 \smns n_3)(1 \smns n_4) \bigr]
	\label{equ: BE}
	.
\end{gather}
Here we introduced the notation $X_j = X(k_j,t)$. $\dE=\ome_1+\ome_2-\ome_3-\ome_4$ is the change in energy and $\dK=k_1+k_2-k_3-k_4$ the change in total momentum. The sum over $m$ allows for Umklapp processes. The matrix element of the Fermi-Hubbard-interaction simply leads to the prefactor $\gamX^2$. The prefactor's time scale is $t_0 = \hbar/\pi J$. For a typical half bandwidth of $J\approx\unit[1]{eV}$ this timescale is $t_0 \approx \unit[0.2]{fs}$.

Note that the Fermi-Dirac distribution $\fd(k)$ makes the collision term vanish, $\Icoll[\fd]=0$ as can be verified easily. This corresponds to the well-known fact that thermal distributions are fixed points of the Boltzmann equation.

Also note that the applicability of the quantum Boltzmann equation relies on fermionic quasiparticle lifetimes of order or larger
than the scattering time. In the one dimensional Hubbard model the appropriate quasiparticles are bosonic (spinons and holons). However,
if their velocities do not differ much one can still use the fermionic quasiparticle picture for not too long times,
which provides the justification for our approach.

\subsection{Dispersion relations}
\label{sec: Dispersion relation}

The \BE describes time evolution due to 2-particle-collisions. The ability of thermalization due to 2-particle-collisions strongly depends on the dispersion relation $\ome(k)$. For instance, if it was quadratic, i.e. $\ome(k)\propto k^2$, momentum conservation $\ddK$ and energy conservation $\ddE$ lead to the following constraints on the 2-particle-collision:
\myEqBegin
	k_1 + k_2 = k_3 + k_4  \ \ \land \ \ k_1^2 + k_2^2 = k_3^2 + k_4^2 
	.
\myEqEnd
This is equivalent to
\myEqBegin
	(k_1=k_3 \ \land \ k_2=k_4) \ \lor \ (k_1=k_4 \ \land \ k_2=k_3)
	.
\myEqEnd
These are two trivial scattering channels. This means that the collision term is zero, i.e. these interaction-channels do not change the \QMD. All dispersion relations permit these trivial channels. While other dispersion relations like nearest-neighbor
hopping, $\ome(k)\propto \cos(2\pi k)$, allow additional scattering processes, the general consensus in the literature is
that in one dimension these additional scattering processes are not sufficient to lead to thermalization. However, building
on the work of F{\"u}rst et al. \cite{Fuerst2012,Fuerst2013}, we will show systematically that adding a next-to-nearest-neighbor
hopping term does indeed lead to thermalization from two particle scattering, contrary to that often stated opinion
in the literature.

\subsection{Linearization}
\label{sec: linearization}

We want to find the relaxation rates. Therefore we linearize the \BE around its thermal distribution, the Fermi-Dirac distribution
\myEqBegin
	\fd(k) := \frac{1}{ \ 1+\exp[\beta (\ome(k)-\mu)] \ }
	.
\myEqEnd
This approximation becomes exact in the limit of small perturbations around the thermal distribution: We will show
that asymptotically a thermal distribution is reached, which therefore provides an a posteriori
justification for the linearization. Hence we obtain the exact time scales describing the late time approach to the
thermal distribution.
The linearization worked out in this subsection follows the scheme described in \citet{Haug1996}.
We start out by introducing a perturbation $\phi$ that we put into the exponent of the Fermi-Dirac distribution:
\myEqBegin
	n(k,t) &= \frac{1}{ \ 1+\exp[\beta(\ome(k)-\mu)-\phi(k,t)] \ }
	\\[3pt] &\approx \fd(k) + \fd(k) \bigl[1-\fd(k)\bigr] \ \phi(k,t)
	.
	\label{equ: linearization}
\myEqEnd
This ansatz has two advantages over the naive scheme $n = \fd+\delta n$.  First one does not have to care so much about the magnitude of $\phi$. In the naive scheme one would have to require
$-\fd(k) \leqslant  \delta n(k,t) \leqslant 1-\fd(k)$.
For us instead it is sufficient to assume
$-1/\bigl[1-\fd(k)\bigr] \leqslant  \phi(k,t) \leqslant 1/\fd(k)$.
The second advantage a much better numerical stability for low temperatures.

Plugging \cref{equ: linearization} into the \BE in \cref{equ: Boltzmann-equation}, we get a rate equation for the perturbation $\phi$ using the stationarity of the Fermi-Dirac distribution, $\Icoll[\fd]=0$:
\begin{gather}
		\dot\phi(k,t) = -\Lc[\phi](k,t)
		,
		\nonumber\\[3pt] \Lc[\phi]_1 = \frac{\BEprefactorLine}{\fd_1 (1\!-\!\fd_1)} \, \intdk3 \, \sddKm \, \ddE 
		\nonumber\\ \qquad \quad \ \ \times (1\!-\!\fd_1) (1\!-\!\fd_2) \fd_3\fd_4 \, \bigl[\phi_1 + \phi_2 - \phi_3 - \phi_4 \bigr]
		\label{equ: Lc}
	.
\end{gather}

Here we have neglected higher order terms in the perturbation $\phi$, so from now on we can work
with the linear operator $\Lc[\phi](k,t)$.

\subsection{Expansion in eigenfunctions}
\label{sec: expansion}

This linear operator $\Lc$ is positive semi-definite and Hermitian with respect to the scalar product
\myEqBegin
	\fscalprod{g}{h} := \int dk \ g(k) \ \fd(k) \bigl[ 1-\fd(k) \bigr] \ h(k)
	,
\myEqEnd
which induces the norm $\fnorm{g}:=\sqrt{\fscalprod gg}$.
The eigenfunctions of $\Lc$ are denoted by $\EF_n$ and the  associated eigenvalues by $\lam_n$. We expand the perturbation $\phi$ in $\EF_n$ using $\fscalprod{\EF_m}{\EF_n}=\fnorm{\EF_n}^2\,\delta_{mn}$ and find
\myEqBegin
	\phi(k,t) = \sum_n A_n \ e^{-\lam_n t} \ \EF_n(k)
	.
\myEqEnd
The coefficients $A_n=\fscalprod{\EF_n}{\phi_0}/\fnorm{\EF_n}^2$ are determined by the initial perturbation $\phi_0(k) = \phi(k,0)$. They measure the contribution of the eigenfunction $\EF_n(k)$ to the perturbation $\phi(k)$. The exponential factor shows that the $\lam_n$ are the rates we are looking for. Due to the fact, that $\Lc$ is positive semi-definite, the eigenvalues are non-negative and we order them by size, $0\leqslant \lam_1 \leqslant \lam_2 \leqslant ...$.

If an eigenvalue is zero, its corresponding contribution $A_n \, \EF_n(k)$ to the perturbation $\phi(k)$ persists for all times. There are two eigenvalues which are always zero:
\myEqBegin
	\lambda_1 = \lambda_2 = 0
	.
	\label{equ: ana0}
\myEqEnd
They correspond to the eigenfunctions $\phi_1(k) = \const$ and $\phi_2(k) = \ome(k)$. A nonzero contribution from these eigenfunctions in our perturbation $\phi$ simply changes the temperature $T$ and the chemical potential $\mu$ according to \cref{equ: linearization}:
\myEqBegin
	\beta_{\text{final}} &= \beta-A_2
	\\ \beta_{\text{final}} \, \mu_{\text{final}} &= \beta\mu - A_1  
\myEqEnd
Therefore these two eigenvalues do not set the thermalization rate: we can eliminate their contributions $A_1$ and $A_2$ by using the correct final temperature and chemical potential in \cref{equ: linearization}. Thus it is the $\nth{3}$ eigenvalue $\lam_3$ which sets the thermalization rate if the initial perturbation has nonzero overlap with the corresponding
eigenvector, $A_3\neq0$.
In general the first eigenvalue $\lambda_n$, $n>2$ with $A_n\neq0$ sets the thermalization rate. We will later see that this is important when approaching half filling because $\lam_3$ shows very different behavior from $\lam_{n>3}$. So its respective eigenfunction $\EF_3(k)$ will be of special interest.

\subsection{Stationary distributions}
\label{sec: stationary qmds}

\newcommand \nqs {n^{\text{S}}}
\newcommand \phiqs {\phi^{\text{S}}}

In the integrable case with nearest-neighbor-hopping only, $\epsX=0$, F{\"u}rst et al. \citet{Fuerst2012} found non-thermal stationary \QMDs. These distributions have the form
\myEqBegin
	\nqs(k) = \frac{1}{ \ 1+\exp[\phiqs(k) + a] \ }
	\label{equ: stationary qmd}
	,
\myEqEnd
where $\phiqs(k)$ is antisymmetric around $k=\pm\tfrac14$ and $a\in \mathds R$ is arbitrary. This means that the $\phiqs(k)$ are stationary perturbations for $\epsX=0$ and the corresponding eigenvalues of the linearized Boltzmann operator vanish.

One can immediately see that all these $\nqs(k)$ form a connected subspace of $L^2((-\tfrac12,\tfrac12])$, i.e. the $\nqs(k)$ can be smoothly transformed into each other. Obviously the thermal \QMDs $1/[1+\exp(\beta (\ome(k)-\mu))]$ also belong to this subspace.
For $\epsX\neq0$ we expect these eigenvectors to change in order $\order(\epsX)$, i.e. $\nqs(k) = \rfrac{1}{1+\exp(\phiqs(k) + a)} + \order(\epsX)$. Furthermore we expect them to become long lived and eventually decay to a thermal \QMD. We denote these quasi-stationary as well as their  associated stationary distributions (for $\epsX=0$) as \QSQMD. They turn out to have the (slightly broken) symmetry
\myEqBegin
	\phiqs \bigl( \pm\tfrac14+k \bigr) = -\phiqs \bigl( \pm\tfrac14-k \bigr) + \order(\epsX)
	\label{equ: Spohn-symm}
	.
\myEqEnd
An arbitrary initial \QMD will relax into a \QSQMD on short time scales. For $\epsX\neq0$ it will then slowly flow to the respective thermal \QMD.

For later reference, Ref.~\citet{Fuerst2012} found that for $\epsX=0$ the only processes that contribute in the \BE fulfill $k_1+k_2 = \pm\tfrac12$.  For small $\epsX$  this constraint becomes 
\myEqBegin
	k_1+k_2 = \pm\tfrac12 + \order(\epsX)
	.
	\label{equ: interaction channels}
\myEqEnd
For larger next-to-nearest-neighbor hopping, $\epsX \geqslant \epsInteractionChannelNumberTransition$, there is an additional interaction channel. In this work we restrict ourselves to $\epsX \leqslant \frac14$, where only one interaction channel has to be considered.


\section{Results}
\label{sec: NumResults}

\subsection{Relaxation rates}

In order to obtain the relaxation rates we first compute a discretized version of the operator $\Lc$. The main difficulty is that a careful interpolation needs to be performed in order to achieve high accuracy. The reason for this is that energy conservation makes it necessary to evaluate the perturbation $\phi$ between grid-points of the discretization. 
Then we diagonalize the discretized operator and perform a finite size scaling analysis of the eigenvalues. A detailed description of the numerics is given in \cref{sec: NumManual}. 

First of all we verify some analytical facts mentioned above.
The first two eigenvalues are at least about $10^{-14}$ times smaller than the largest eigenvalue. 
As expected from the discussion in \cref{sec: expansion}, the corresponding eigenfunctions are superpositions of the constant function and the dispersion relation.
%
\cref{fig: 3rd EVec,fig: Evecs 4 5 6}
show eigenfunctions of $\Lc$ correspoding to low lying (nonvanishing) eigenvalues. They are antisymmetric around $k=\pm\frac14$ up to a correction of $\order(\epsX)$. So they have the correct symmetry as given by \cref{equ: Spohn-symm} to be a perturbation $\phiqs(k)$ corresponding to a \QSQMD.

\putFigureEigenfunctionIII

\putFigureEigenfunctionsIVToVI

\putFigureEigenvalueIIIEpsDependence

\putFigureEigenvalueIIITempertureDependence

The two double logarithmic plots in  \cref{fig: JNNN dependence} show the lowest non-zero eigenvalues $\lam_{n\geqslant3}$ for  $\beta=0.1$ and $\beta=10.0$. Plotted are the eigenvalues after extrapolating to an infinitesimal discretization grid of the linearized Boltzmann operator. The lines are fits with a quadratic $\epsX$-dependence. For the fits we only used points in the $\epsX^2$-regime of the eigenvalues. These fits show that for sufficiently small $\epsX$ every low lying eigenvalue has a quadratic $\epsX^2$ dependence on the relative strength to the next-to-nearest-neighbor hopping. \cref{fig: JNNN dependence} also shows that the linear fits are an upper bound to their respective eigenvalue. 

From this
data we can already conclude that nearest-neighbor hopping plus nonvanishing
next-to-nearest-neighbor hopping is indeed sufficient to achieve thermalization via two particle scattering in one dimension.
The thermalization rates are quadratic in the relative strength of the next-to-nearest-neighbor hopping~$\epsX$
(for small~$\epsX$).

Similar data like in \cref{fig: JNNN dependence} was obtained for chemical potentials $\mu\in\{0,\pm0.1,\pm0.9\}$ and for the inverse final temperature ranging from $\beta=0.01$ to $\beta=30$. In all cases the eigenvalues are proportional to $\epsX^2$ for sufficiently small $\epsX$, so the lowest non-zero eigenvalues obey
\myEqBegin
	\lam_n(\JNN,\gamma,\epsX,\beta,\mu) 
		= J \ \tlam_n(\beta,\mu) \ \gamX^2 \ \epsX^{2}
		\label{equ: relaxation rates 1}
\myEqEnd
for $\gamma$ and $\epsX$ sufficiently small. Here $\tlam_n(\beta,\mu)$ are the proportionality factors which we will denote ``rescaled eigenvalues''. One can also show analytically that there are no terms proportional to $\epsX$.

\putFigureUmklapp

\cref{fig: beta dependence} shows  three plots depicting the rescaled eigenvalues $\tlam_{n \geqslant 3 }(\beta,\mu)$ as functions of $\beta$. The first non-zero eigenvalue $\tilde\lambda_3(\beta)$ decreases
exponentially as a function of inverse temperature for all fillings. The higher eigenvalues ($n\geq 4$) decay
exponentially away from half filling, but are asymptotically constant for $T\rightarrow 0$ at half filling.
The straight lines are fits to exponential behavior which we can parametrize as
\myEqBegin
	\lam_n \propto e^{-\Gamma_n \beta |\mu|} \quad \text{ for } n \geqslant 4 
	\label{equ: relaxation rates 2}
	.
\myEqEnd
This behavior can be explained by Umklapp processes (see also Ref.~\cite{Rosch2000}). At half filling the situation is shown in \cref{fig: Umklapp}. Consider for example two particles with momenta $k_{1,2}\approx \tfrac14$. They can scatter into $k_3\approx-\tfrac14$ and $k_4\approx\tfrac34$. The latter becomes $k_4\approx\tfrac34-1 = -\tfrac14$ by subtraction of the reciprocal lattice vector. This process becomes less likely away from half filling. The Fermi points shift apart (or together) when changing the filling and Umklapp processes like the one depicted in \cref{fig: Umklapp} cannot happen any more. Therefore  Umklapp processes are most effective at half filling for low final temperatures.

Therefore generically thermalization becomes exponentially slow as a function of the inverse final temperature (equivalently: 
the inverse initial excitation energy) away from half filling. At half filling only the eigenvector $\EF_3(k)$ shows this
behavior, so at half filling there is no exponential slowdown of thermalization as a function of inverse final temperature
if the initial perturbation does not couple to this eigenvector.
Therefore this eigenvalue and its eigenfunction $\EF_3(k)$ are of special interest. The next sections will deal with understanding this eigenfunction $\EF_3(k)$.

\subsection{Structure of $\boldsymbol{\EF_3(k)}$}
\label{sec: approx EF}

The values of the initial perturbation $\phi(k)$ around $k\!=\!0$ and $k\!=\!\pm\frac12$ are not important for small temperatures since they are suppressed by the factor $\fd(k)[1\!-\!\fd(k)]$, see \cref{equ: linearization}.
So based on \cref{fig: 3rd EVec}
a good approximation of the third eigenfunction is given by
\myEqBegin
	\EFapprox(k) := k-\tfrac14 \sgn(k)
	.
	\label{equ: EFapprox}
\myEqEnd
This is shown in \cref{fig: EFapprox}. One can see that  $1-\fscalprod{\EF_3}{\EFapprox} / \fnorm{\EF_3}\fnorm{\EFapprox}$ approaches zero in the limit $T\rightarrow0$, which means that indeed $\EFapprox(k)$ is a good approximation.
Using this result one can show \cref{sec: Scattering processes in EF3} that  Umklapp and forward scattering do not contribute in  $\Lc[\EFapprox](k_1)$, only backscattering can occur. But backscattering is limited to $k_1$ being in the vicinity of $0$ or $\pm\frac12$, which leads to the exponential suppression of $\lam_3$ as a function of inverse temperature
\cref{sec: Scattering processes in EF3}.

\newcommand\putFigureOverlapWithEFapprox{
\begin{figure}
\begin{center}
\includegraphics[width=\halfWidth
	]{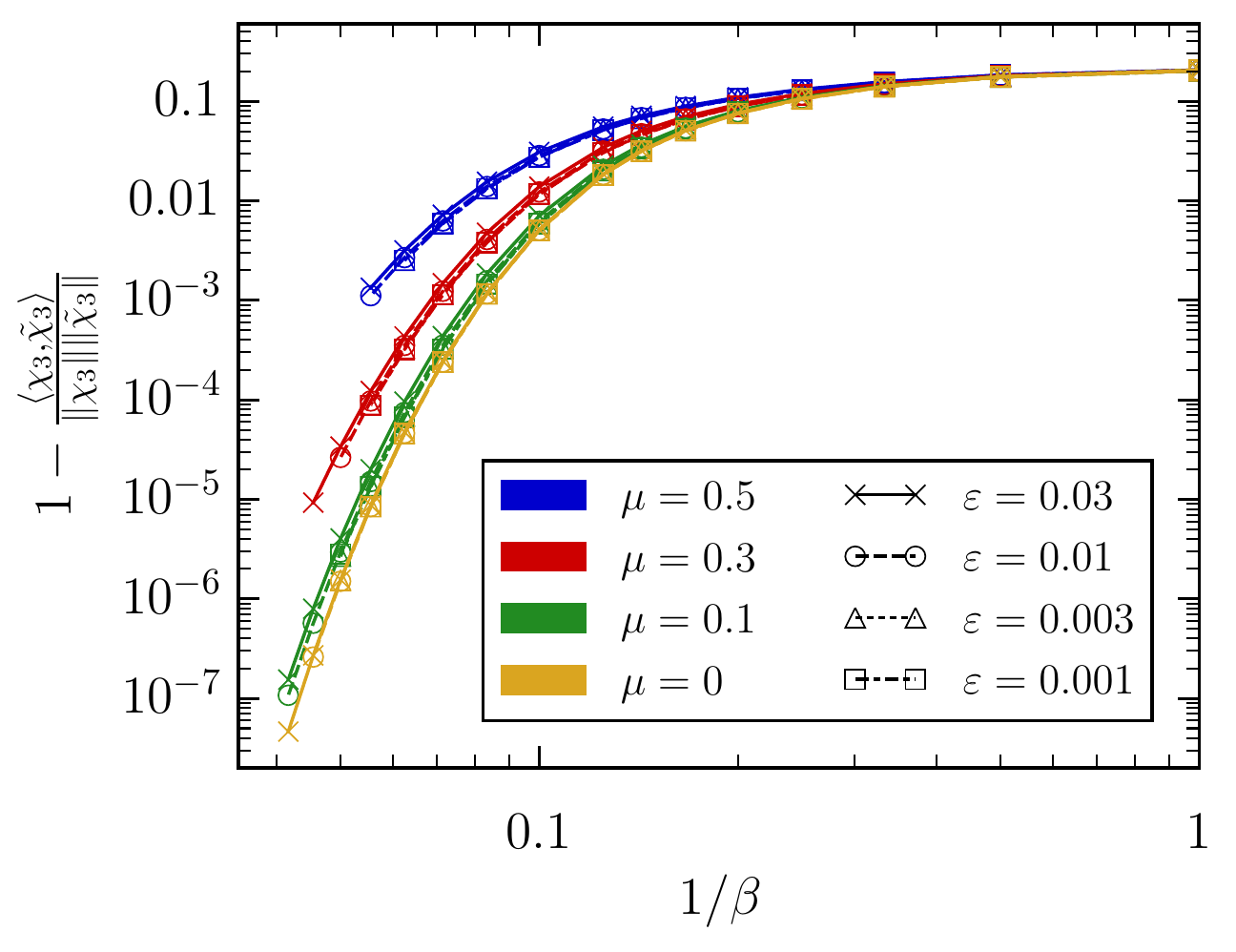}
\caption{\label{fig: EFapprox} This plot shows that $\EFapprox(k)$ becomes a very good approximation for $\EF_3(k)$ at low temperatures $T$.}
\end{center}
\end{figure}
}
\putFigureOverlapWithEFapprox

In the following we consider two operators which have a connection to $\EF_3(k)$. The first operator, $\Iiii$, comes from the approximate eigenfunction $\EFapprox(k)$.
The other operator will turn out to be the total energy current operator $\JE$.

Let us first construct operators in state space using the eigenfunctions $\operatorEF_n(k)$ of $\Lc$:
\myEqBegin
	\operatorEF_n := \int dk \ \EF_n(k) \ \n(k)
	.
\myEqEnd
Their expectation values are
\myEqBegin
	\expectShort{ \operatorEF_n }
	&= \int dk \ \EF_n(k) \ n(k,t)
	\overset{{(\ref{equ: linearization})}}= C_n + \fscalprod{\EF_n}{\phi} +\order(\phi^2)
	\\& \approx C_n + A_n \ e^{-\lam_n t} \ \fnorm{\EF_n}^2
	\label{equ: expectation of In}
	,
\myEqEnd
where $C_n:= \int dk \ \EF_n(k) \ \fd(k)= \const$.
Since   $\expectShort[0]{ \operatorEF_n }-C_n$ is proportional to $A_n$ this indicates if there is a slowly decaying perturbation in the system. 

An operator $\hat Q_n$ can be used for the construction of a generalized Gibbs ensemble (GGE) that describes the long time limit of the QMD provided that $\lambda_n = 0$. For the integrable case, $\epsX = 0$, there are infinitely many zero eigenvalues and therefore an infinite number of conserved charges $Q_n$ that enter such a GGE.

Now we consider specifically $n=3$. As discussed above, the eigenvalue $\lam_3$ is very small for low final temperatures. So $\Iiii$  may be used for the construction of a GGE (see \cref{sec: GGE}), that describes the \QMD on time scales $1/\lam_4 \ll t \ll 1/\lam_3$.
%
In order to give the  the operator $\Iiii$ a physical meaning, we replace $\EF_3(k)$ with its approximation $\EFapprox(k)=k-\frac14 \sgn(k)$, \cref{equ: EFapprox}:
\myEqBegin
	\Iiii \approx C \int dk \ \EFapprox(k) \ \n(k) = C \bigl[\ktot - \tfrac14 (\NR - \NL) \bigr]
\myEqEnd
with the constant $C=\fnorm{\EF_3}/\fnorm{\EFapprox}$.
The momentum term of $\EFapprox(k)$ leads to the total momentum operator $\ktot$, the signum function results in the operator $\NR-\NL$. Here $\NL$ counts the particle number on the left side of the Brioullin zone, and $\NR$ the particle number on the right side.
The reason for the long living expectation value of $\Iiii$ is the same as for the approximate stationarity of $\EFapprox(k)$: Umklapp processes are ineffective even at half filling.

\paragraph*{\textbf{Long living currents:}}
On times $1/\lam_4 \ll t \ll 1/\lam_3$ the perturbation has decayed to $\phi(k) \approx e^{-\lam_3 t} A_3 \EF_3(k)$ and its exponential factor $e^{-\lam_3 t}$ is still approximately one. \cref{fig: 3rd EVec} shows that $\EF_3(k)$ is asymmetric around $k=0$. So the corresponding momentum distribution $n(k,t)$ is also asymmetric corresponding to nonvanishing currents. The operators of the total particle and energy current are
\myEqBegin
	\JNE = \sum_l \jNEl + \order(U) = \int \! dk \ \jNE(k) \, \n(k) + \order(U)
	,
\myEqEnd
respectively.
The local current operators $\jNl$ and $\jEl$ fulfill a discretized version of the continuity equation. $\jN(k):= 2J \, \groupvel$ is simply the group velocity and $\jE(k):=(2J)^2 \, \groupvel \, \disprel$ also includes the dispersion relation. Following the scheme in \cref{equ: expectation of In}, the expectation values evaluate to
\myEqBegin
	\langle \JNE \rangle \approx e^{-\lam_3 t} \, \frac{\fscalprod{\phi_0}{\EF_3}\fscalprod{\EF_3}{\jNE}}{\fnorm{\EF_3}^2}
	.
	\label{equ: long lasting current}
\myEqEnd
One can show numerically and analytically that $\fscalprod{\jNE}{\EF_3}$ is  not zero.
Notice that the right hand side of \cref{equ: long lasting current} is independent of system size, as is the left hand side.
Therefore there are long living currents if one creates an initial perturbation that fulfills $\fscalprod{\phi_0}{\EF_3} \neq 0$.

\section{Conclusions}
\label{sec: Conclusions}

%
We have investigated the long time behavior of the quasi-momentum distribution of a one-dimensional fermionic Hubbard model
with additional next-to-nearest-neighbor hopping~$J'$. Using a linearized Boltzmann equation we could systematically 
verify that thermalization occurs from two particle scattering processes if $J'$ is nonzero
\cite{Fuerst2012, Fuerst2013}. It is often stated that a Boltzmann equation with two particle scattering is
ineffective in one dimension because of simultaneous energy and momentum conservation. Following
F\"urst et al. \cite{Fuerst2012, Fuerst2013} we have therefore verified that this statement is incorrect for a
band dispersion described by nearest neighbor plus next-to-nearest-neighbor hopping.

Away from half filling the relaxation rates are quadratic in the relative strength $\epsX$ of the next-to-nearest-neighbor hopping
and are exponentially suppressed as a function of inverse final temperature. This implies that thermalization occurs,
but on an exponentially increasing time scale for low excitation energy of the initial perturbation. 
At half filling this picture is different since Umklapp processes play an important role even at low temperatures:
at half filling the only perturbation with such an exponential slowdown corresponds to the current, all other perturbations
decay with a constant rate (still proportional to $\epsX^2$) in the low temperature limit.
Notice the difference to the behavior predicted by the \BE in higher dimensions where the smallest relaxation rates are proportional to $T^2$.\citep{Kabanov2008} In a pump-probe experiment this would translate into a much stronger (exponential)
dependence of the thermalization time scale on fluence in one dimension.

\section{Acknowledgments}

We thank H.~Spohn, A.~Mitra, G.~Mussardo, M.~Medvedyeva, I.~Homrighausen, D.~Fioretto, R.~H\"artle, and S.~R.~Manmana for helpful discussions. This work was supported by the SFB1073 (project B03) of the Deutsche Forschungsgemeinschaft (DFG).



\begin{appendix}

\section{Stationary distributions of the standard FHM}
\label{sec: Stationarity of nqs}

For $\epsX=0$ there are many stationary distributions $\nqs(k)$ \citet{Fuerst2012}, see \cref{equ: stationary qmd}.
Momentum and energy conservation lead to $k_2=\tfrac12-k_1$ and $k_4=\tfrac12-k_3$. Therefore 
the gain minus loss term in the Boltzmann equation vanishes, $(1 \smns n_1)(1 \smns n_2)n_3n_4 - n_1n_2(1 \smns n_3)(1 \smns n_4)=0$  because
\myEqBegin
	&(1 \smns \nqs_1)(1 \smns \nqs_2) \nqs_3 \nqs_4
	\\ & \ \ = \frac1{e^{-\phiqs_1+a}+1} \ \frac1{e^{\phiqs_1+a}+1}
		\ \frac1{e^{\phiqs_3-a}+1} \ \frac1{e^{-\phiqs_3-a}+1}
	\\ & \ \ = \nqs_1 \nqs_2 (1 \smns \nqs_3)(1 \smns \nqs_4)
		\, \underbrace{ e^{\phiqs_1-a}  e^{\phiqs_1-a}  e^{\phiqs_3+a}  e^{-\phiqs_3+a} }_{\displaystyle =1}
	.
\myEqEnd
This yields
\myEqBegin
	\Icoll[\nqs]=0
	.
\myEqEnd
and $\nqs(k)$ is indeed a stationary solution.

\section{Numerics}
\label{sec: NumManual}

\putFigureEigenvaluesTest

\putFigureTL

To get the thermalization rates we need to find the eigenvalues of $\Lc$. Therefore we discretize the operator, diagonalize it and extrapolate its spectrum to the continuum limit.

We discretize $\Lc$ by discretizing momentum space:
\myEqBegin
	k_1,k_3 \in \bigl\{-\tfrac12\!+\!\tfrac1{\Nk},...,\tfrac12\!-\!\tfrac1{\Nk},\tfrac12\bigr\}=:\Mk
\myEqEnd
The discretized operator is denoted by the matrix $\Lk_{k_1,q}$, the discretized perturbation by the vector $\phiv[q](t)$:
\myEqBegin
	\sum_{q\in \Mk} \Lk_{k_1,q} \; \phiv[q](t) \, \xrightarrow{\Nk\to\infty} \, \Lc[\phi](k_1,t)
	.
\myEqEnd
Then we need to obtain the non-trivial solutions of  $\dE(k_1,k_2,k_3)\equiv \ome(k_1)+\ome(k_2)-\ome(k_3)-\ome(k_4=k_1+k_2-k_3)=0$.
It can be shown that for every $(k_1,k_3)\in \Mk^2$ there is exactly one non-trivial solution $\tk_2\equiv\tkf$ as long as $0 \leqslant \epsX \leqslant \epsInteractionChannelNumberTransition$. This is because the dispersion relation $\ome(k)$ is symmetric around zero, monotonic in the interval $(0,1/2)$, and periodic.
We numerically calculate the solutions $\tkf$ by a Newton-Raphson procedure. In general these solutions are not on the grid  $\Mk$. But considering the formula for $\Lc$, \cref{equ: Lc}, we realize that the perturbation $\phi$ has to be evaluated at $\tk_2$. This can only be done with an interpolation of the discretized perturbation $\phiv[q]$.

This is why we have to use an interpolation scheme to evaluate $\ddE$. A naive interpolation, like using step functions or linear functions, leads to large errors. So we use
\myEqBegin
	&\intdk2 \ \ddE \ g_{k_1,k_3}(k_2) 
	\\ &= \int dk_3 \, \sum_{\tk_2} \, \frac{1}{|d\dE/d\tk_2|} \ g_{k_1,k_3}(\tk_2)
	\\ &= \lim_{\Nk\to\infty} \frac1{\Nk} \sum_{k_3\in\Mk} \sum_{i=1}^{\Ni} \frac{C_i(\tk_2)}{|d\dE/d\tk_2|} \ g_{k_1,k_3}\bigl(p_i(\tk_2)\bigr)
	, \label{equ: ddE}
\myEqEnd
where $\Ni$ is the number of interpolation momenta, $p_i\!\in\!\Mk$ are the nearest discretized momenta next to $\tk_2$ and $C_i$ the respective weights. These weights are calculated such that $g_{k_1,k_3}(k_2)$ is approximated in the best possible way. In our case there is exactly one solution for a given tuple $(k_1,k_3)$. This is why the sum over $\tk_2$ is dropped in the last line of Eq. \cref{equ: ddE}.

Using our interpolation scheme for $\Lk_{k_1,q}$ we obtain
\begin{gather}
	\Lk_{k_1,q} \!=\! \frac{\BEprefactorEnum}{\BEprefactorDenom \Nk} \! \sum_{k_3\in\Mk} \sum_{i=1}^{\Ni} \frac{1\!-\!\fdm\bigl(p_i(\tk_2)\bigr)}{\fd_1} \fd_3 \fdm\bigl(k_1\!-\!k_3\!+\!p_i(\tk_2)\bigr) 
	\nonumber \\ \quad \ \times \Bigl[ \delta_{k_1,q} \!+\! \delta_{p_i,q} \!-\! \delta_{k_3,q} \!-\! \delta_{k_1-k_3+p_i,q} \Bigr] \ \frac{C_i(\tk_2)}{|d\dE/d\tk_2|}
\end{gather}
For our purposes we found that $\Ni=9$ is a reasonable choice. The speed of the numerical evaluation is still fast enough while providing high precision. This is shown in \cref{fig: N-point-interpolation}. There we plotted the first three eigenvalues. $\lam_1$ is fluctuating around $10^{-16}$. The reason is the simple nature of its corresponding eigenfunction $\EF_1=\const$. It is an exact eigenfunction for any discretization in our numerics due to the fact that $\phi_1+\phi_2-\phi_3-\phi_4 \equiv 0 \, \forall k_{1,2,3,4}$ trivially. So $\lam_1$ measures the 
\emph{resolution of zero}. The eigenfunction $\EF_2$, however, is curved and therefore the numerical eigenfunction is influenced by the interpolation scheme. Thus $\lam_2$ measures the 
\emph{numerical precision including the interpolation scheme}. Additionally we plotted $\lam_3$ in \cref{fig: N-point-interpolation} as a reference for the other eigenvalues. One can see that from $\Ni=3$ on it does not change any more and becomes independent of our interpolation scheme.


The eigenvalues of $\Lk$ are found by exact diagonalization. These are then extrapolated to $\Nk\rightarrow\infty$ to obtain the eigenvalues of $\Lk_{k_1,q}$, see \cref{fig: TL}. Figs.~\ref{fig: TL}A and B show that for lower $\epsX$ oscillations are visible for the lower eigenvalues. But for higher $\beta$ (like in C) these oscillations disappear. However, $\lam_{50}$ always shows a strong exponential decay towards $\Nk\to\infty$. So one has to cover a variety of extrapolations. Therefore we use a script, which extrapolates our data for several $\epsX$, $\beta$ and $\mu$. It fits a constant function, an exponential function and the function $x \mapsto a \exp(b x) \sin(c x)$ to the data. Then the script chooses the one with the smallest error (straight lines). The value at $1\L=0$ as well as the error are used to create the plots in \cref{fig: JNNN dependence}.

\putFigurekjs

\section{Scattering processes for \texorpdfstring{$\boldsymbol{\EF_3}$}{chi_3}}
\label{sec: Scattering processes in EF3}

In \cref{sec: approx EF} we consider $\Lc[\EFapprox](k_1)$ to explain the smallness of the eigenvalue $\lam_3$ corresponding to the eigenfunction $\EF_3$. In order to do that we need two ingredients. Note that both apply to all eigenfunctions. The first one is the fact that the mean momentum $\frac12(k_1+k_2)=\frac12(k_3+k_4)$ lies in an $\epsX$-region around $\tfrac14$, see \cref{equ: interaction channels}.
The second ingredient is the structure of $\Lc$, which contains all the Fermi functions:
\myEqBegin
	\frac{1 \smns \fd_2}{\fd_1}\fd_3\fd_4 &=
		\frac{1}{\fd_1}(1 \smns \fd_1 \spls \fd_1)(1 \smns \fd_2)\fd_3\fd_4
	\\ &\overset{\mathclap{\dE = 0} \quad}= \ \ \   \fd_2(1 \smns \fd_3)(1 \smns \fd_4) + (1 \smns \fd_2)\fd_3\fd_4
	\label{equ: ingredient 2}
	.
\myEqEnd
For $\mu$ away from full or empty filling both \cref{equ: interaction channels,equ: ingredient 2} make $k_3$ and $k_4$ lie on the same side of the Brioullin zone. This means that either both are positive or both are negative. Since we only explain the linear regions of $\EF_3(k_1)$, $k_1$ is away from $0$ and $\pm\tfrac12$. This and \cref{equ: interaction channels} make also $k_1$ and $k_2$ lie on the same side of the Brioullin zone. Thus in $\Lc[\EFapprox](k_1)$ with $k_1$ around $\pm \frac14$ there is no backward scattering like \cref{fig: LLRR}B. There is only forward and Umklapp scattering like illustrated in \cref{fig: LLRR}A and C, respectively.

Now we use the linear form of $\EFapprox(k_1)$ and consider the factor
\myEqBegin
	\EFapprox(k_1) + \EFapprox(k_2) - \EFapprox(k_3) - \EFapprox(k_4)
	\label{equ: factor}
\myEqEnd
in $\Lc[\EFapprox](k_1)$. The approximate eigenfunction $\EFapprox(k)=k-\frac14 \sgn(k)$, see \cref{equ: EFapprox}, consists of two terms. These are the momentum or identity function $\id(k)=k$ and the signum function $\sgn(k)=\pm1$. We calculate \cref{equ: factor} for both functions separately.

First we consider forward scattering, i.e. all $k_j$ have the same sign, see e.g. \cref{fig: LLRR}A. Plugging the momentum function into \cref{equ: factor} gives $k_1+k_2-k_3-k_4 = 0$ because forward scattering processes obey momentum conservation. Plugging the signum function into \cref{equ: factor} gives zero, too. So forward scattering does not contribute to $\Lc[\EFapprox](k_1)$.

Now we consider Umklapp scattering, see e.g. \cref{fig: LLRR}C. There momentum is not conserved. Thus plugging the momentum function into \cref{equ: factor} we obtain  the momentum change $k_1+k_2-k_3-k_4 = \pm 1$. It is compensated by the signum function plugged into \cref{equ: factor}: $\sgn(k_1)+\sgn(k_2)-\sgn(k_3)-\sgn(k_4) = \pm 4$. This has the same sign as the momentum change, which leads to cancellation in \cref{equ: factor}. This explains the signum function's prefactor of $-\frac14$ in the approximate eigenfunction. So Umklapp scattering does not contribute to $\Lc[\EFapprox](k_1)$, similar to forward scattering. 


\newcommand \dtheta{\vartheta}
\newcommand \btheta{\bar{\dtheta}}
\newcommand \phinorm{\mathcal{N}}

\section{GGE for \texorpdfstring{$\boldsymbol{\lam_3}$}{lambda_3}}
\label{sec: GGE}
In \cref{sec: linearization} we learned that the first non-trivial eigenvalue is $\lam_3$. In \cref{fig: 3rd EVec} we saw that for $\beta\to\infty$ the  associated eigenfunction $\phi_3(k)$ approaches $\EFapprox(k)=k-\tfrac14 \sgn(k)$. In the following we derive the  respective (approximately) conserved quantity. The generalized Gibbs ensemble (GGE) is
\myEqBegin
	Z_{\text{GGE}} &= \tr e^{-\beta \Ham + \beta \mu \hat{N} - \beta_3 \Iiii}
	.
\myEqEnd
The conserved quantity has the form
\myEqBegin
	\quad \Iiii = \sum_k \alpha(k) \ \n(k)
	.
\myEqEnd
We find
\myEqBegin
	\sum_k \nGGE(k) &\!=\! \frac{\partial}{\partial(\beta\mu)} \ln \ZGGE
	\!=\! \sum_k \frac{1}{1\!+\!e^{\beta (\omega(k)-\mu) + \beta_3 \alpha(k)}}
\myEqEnd
analogous to the standard text book derivation of the Fermi-Dirac-distribution. Comparing this to \cref{equ: linearization}, where we defined our linearization of $n(k,t)$, we identify $\alpha(k) \approx \EFapprox(k)$. Therefore
\myEqBegin
	\Iiii &\approx \sum_k k \, \n(k) - \tfrac14 \sum_{\mathclap{k>0}} \n(k) + \tfrac14 \sum_{\mathclap{k<0}} \n(k)
	\\ &=\ktot - \tfrac14 (\NR - \NL)
	.
\myEqEnd
Note that the factor $\tfrac14$ does not depend on $\kF$. It is due to the fact that $\phi(k)$ has to be antisymmetric around $k=\pm\tfrac14+\order(\epsX)$ for a quasi-stationary state, see \cref{sec: stationary qmds}.

\end{appendix}



\bibliography{paper_Biebl_Kehrein}

\end{document}